\documentclass[12pt]{article}
\usepackage[bookmarks=true]{hyperref}
\usepackage{sbc-template}

\usepackage{graphicx,url}

\usepackage[utf8]{inputenc}

\usepackage{booktabs}
\usepackage{multirow}
\usepackage{color}

\usepackage[inline]{enumitem}

\usepackage{empheq}
\usepackage{multicol}

\usepackage[table]{xcolor}

\usepackage{listings}
\usepackage{inconsolata}
\definecolor{pblue}{rgb}{0.13,0.13,1}
\definecolor{pgreen}{rgb}{0,0.5,0}
\definecolor{pred}{rgb}{0.9,0,0}
\usepackage{amssymb,amsmath}
\newcommand{\keepcomment}{1} %@Fer: 1 to show the comments, 0 to hide them

\newcommand{\bnote}[3]{
\ifnum\keepcomment=1
	\fbox{\bfseries\sffamily\scriptsize#3{#1}}
	{\sf\small$\blacktriangleright$\textit{#2}$\blacktriangleleft$}
\else
\fi
}

\newcommand{\ToolName}{PPD}

\title{Towards an automated approach for bug fix pattern detection}

\author{Fernanda Madeiral\inst{1}, Thomas Durieux\inst{2}, Victor Sobreira\inst{1}, Marcelo Maia\inst{1}}

\address{
Federal University of Uberl\^andia, Brazil
\nextinstitute
INRIA \& University of Lille, France
\email{fernanda.madeiral@ufu.br, thomas.durieux@inria.fr, \{victor, marcelo.maia\}@ufu.br}
}

\begin{document} 

\maketitle

\begin{abstract}
%Software bug-related research fields such as automatic program repair rely on bug datasets to assess the performance of proposed approaches.
%Selecting bugs and performing advanced analysis when evaluating automatic program repair tools are not trivial tasks since detailed information on the bugs and their patches (bug fixes) are usually not available.
The characterization of bug datasets is essential to support the evaluation of automatic program repair tools.
In a previous work, we manually studied almost 400 human-written patches (bug fixes) from the Defects4J dataset and annotated them with properties, such as repair patterns. However, manually finding these patterns in different datasets is tedious and time-consuming.
To address this activity, we designed and implemented \ToolName, a detector of repair patterns in patches, which performs source code change analysis at abstract-syntax tree level. In this paper, we report on \ToolName~and its evaluation on Defects4J, where we compare the results from the automated detection with the results from the previous manual analysis. We found that \ToolName~has overall precision of 91\% and overall recall of 92\%, and we conclude that \ToolName~has the potential to detect as many repair patterns as human manual analysis.
\end{abstract}

\section{Introduction}
\vspace*{-2pt}

Automatic program repair is a recent  research field where approaches have been proposed to fix software bugs automatically, without human intervention \cite{Monperrus2018}.
In automatic program repair, empirical evaluation is conducted by running repair tools on \textit{known bugs} to measure the repairability potential. These known bugs are available on bug datasets, e.g., Defects4J~\cite{Just2014}.

Building datasets of bugs is a challenging task. Despite the effort made by authors of bug datasets, such datasets generally do not include detailed information on the bugs and their patches (bug fixes), so a fair and advanced evaluation of repair tools becomes harder. We highlight two  tasks that make the evaluation of repair tools more robust:

\begin{itemize}[leftmargin=1.2em]
\item \textit{Selection of bugs} is used to filter out bugs that do not belong to the bug class which the repair tool under evaluation targets.
For instance, NPEFix~\cite{Durieux2017-npefix} is a tool specialized in null pointer exception fixes, and it will probably fail on bugs that were not fixed by a human with a null pointer checking. So, a coherent and fair analysis would include only bugs within the target bug class(es) of the respective tools.

\item \textit{Correlation analysis} is an advanced analysis of the results produced by a repair tool, making possible to derive conclusions such as ``the repair tool performs well on bugs having the property X''. This kind of analysis requires a characterization of all bugs available in the used dataset. 
\end{itemize}

To support these two tasks, in a previous work~\cite{Sobreira2018defects4J-dissection}, we \textit{manually} analyzed the patches of the Defects4J dataset \cite{Just2014}, which is a widely used dataset in automatic program repair field. As a result, we delivered a taxonomy of properties and the annotation of Defects4J patches according to such taxonomy. Despite the value of manual work, analyzing patches to calculate or to find properties when characterizing patches from different datasets is tedious and time-consuming. Nevertheless, the already built taxonomy is a useful resource to guide the automation of patch analysis.

In this paper, we present \ToolName~(\underline{P}atch \underline{P}attern \underline{D}etector), a detector of \textit{repair patterns} in patches, one of the existing types of property in our previous taxonomy. Repair patterns are recurring  abstract structures in patches. For instance, a patch that affects only one line in the buggy code is an instance of the pattern \textit{Single Line}, while a patch that adds an entire conditional block is an instance of the pattern \textit{Conditional Block Addition}.

\ToolName~analyzes a given patch by first retrieving edit scripts at Abstract-Syntax Tree (AST) level from the {\it diff} between the buggy and patched code using GumTree~\cite{Falleri2014}. Then, PPD searches for instances of patterns by analyzing the AST  nodes of the {\it diff} using Spoon~\cite{Pawlak2015}, a peer-reviewed library to analyze Java source code. We evaluated \ToolName~and found that it has the potential of detecting the nine repair pattern groups from Sobreira et al. (2018). \ToolName~can support automatic repair researchers on selecting bugs from bug datasets and performing correlation analysis between repaired bugs and their properties. Moreover, \ToolName~can be useful in comparisons between different datasets of bugs. By discovering the constitution of bug datasets, it is possible to study the balance between them and also the flaws.

To sum up, the main contributions of this paper are 1) a tool to detect repair patterns from patches written in Java, which is publicly available, and 2) the detection of new 90 instances of the patterns on Defects4J.

%@Fer: the following is useless for short paper. We have more important things to say ;)
%The remainder of this paper is organized as follows. Section~\ref{sec:background} presents the taxonomy of repair patterns. Section~\ref{sec:tool} presents \ToolName. Section~\ref{sec:evaluation} presents the evaluation of \ToolName, containing the used method and the obtained results, as well as a discussion on the results. Section~\ref{sec:threats} discusses threats to validity. Section~\ref{sec:related-works} presents the related work, and Section~\ref{sec:conclusion} presents the conclusions.

\section{Taxonomy of Repair Patterns}\label{sec:background}
\vspace*{-2pt}

Previously, we delivered a taxonomy of repair patterns containing nine groups and 25 patterns in total~\cite{Sobreira2018defects4J-dissection}. \ToolName~implements the detection of all these patterns. In this section, we briefly define the nine pattern groups. Additionally, we refer in this paper to patches from Defects4J (using a simple notation with project name followed by bug id) to provide examples with external links for patch visualization, e.g., \href{http://program-repair.org/defects4j-dissection/#!/bug/Chart/1}{Chart-1}\footnote{For paper printed version: all links on Defects4J patches can be built by inserting two parameters in \url{http://program-repair.org/defects4j-dissection/\#!/bug/<project_name>/<bug_id>}. Example: \href{http://program-repair.org/defects4j-dissection/\#!/bug/Chart/1}{Chart-1} contains the link \url{http://program-repair.org/defects4j-dissection/\#!/bug/Chart/1}}.

\vspace*{-3pt}
\noindent\textit{Conditional Block} involves the addition or removal of conditional blocks (e.g., \href{http://program-repair.org/defects4j-dissection/#!/bug/Lang/45}{Lang-45}).

\vspace*{-3pt}
\noindent\textit{Expression Fix} involves actions on logic (\href{http://program-repair.org/defects4j-dissection/#!/bug/Chart/1}{Chart-1}) or arithmetic (\href{http://program-repair.org/defects4j-dissection/#!/bug/Math/80}{Math-80}) expressions.

\vspace*{-3pt}
\noindent\textit{Wraps/Unwraps} consists of (un)wrapping existing code with/from high-level structures such as try-catch blocks (\href{http://program-repair.org/defects4j-dissection/#!/bug/Closure/83}{Closure-83}) and low-level ones such as method calls (\href{http://program-repair.org/defects4j-dissection/#!/bug/Chart/10}{Chart-10}).

\vspace*{-3pt}
\noindent\textit{Single Line} is dedicated to patches affecting one single line or statement (\href{http://program-repair.org/defects4j-dissection/#!/bug/Closure/55}{Closure-55}).

\vspace*{-3pt}
\noindent\textit{Wrong Reference} occurs when the code references a wrong variable (e.g., \href{http://program-repair.org/defects4j-dissection/#!/bug/Chart/11}{Chart-11}) or method call (e.g., \href{http://program-repair.org/defects4j-dissection/#!/bug/Closure/10}{Closure-10}) instead of another one.

\vspace*{-3pt}
\noindent\textit{Missing Null-Check} is related to the addition of a conditional expression or the expansion of an existing one with a null-check that was missing in the code (e.g., \href{http://program-repair.org/defects4j-dissection/#!/bug/Chart/15}{Chart-15}).

\vspace*{-3pt}
\noindent\textit{Copy/Paste} is the application of the same change to different points in the code (\href{http://program-repair.org/defects4j-dissection/#!/bug/Chart/19}{Chart-19}).

\vspace*{-3pt}
\noindent\textit{Constant Change} involves changes in literals or constant variables (e.g., \href{http://program-repair.org/defects4j-dissection/#!/bug/Closure/65}{Closure-65}).

\vspace*{-3pt}
\noindent\textit{Code Moving} involves moving code statements or statement blocks around, without extra changes to these statements (e.g., \href{http://program-repair.org/defects4j-dissection/#!/bug/Closure/117}{Closure-117}).
%In other words, the moved statements are deleted in one location and added in another location in the code.
%It may consist of single statement as in Closure-\href{http://program-repair.org/defects4j-dissection/#!/bug/Closure/13}{13}, or multiple statements as in Closure-\href{http://program-repair.org/defects4j-dissection/#!/bug/Closure/117}{117}.
%We do not consider a code as being moved when it is a side effect from the occurrence of other patterns, for example \textit{Wraps-with}.

\section{\ToolName: a Detector of Bug Fix Patterns}\label{sec:tool}
\vspace*{-2pt}

The detection of repair patterns in patches falls in source code change analysis task.
Analyzing source code changes can be performed at different levels of granularity such as file level, line level, and AST level. Our approach is at the AST level, and it consists of two main tasks to detect repair patterns in a given patch:

\noindent\textit{Retrieval of the AST diff:} Given as input the buggy version of the program and the patch file (\textit{diff} file), \ToolName~retrieves the AST \textit{diff} between the buggy and patched code (also known as \textit{edit scripts}) using the GumTree algorithm~\cite{Falleri2014}. There are different implementations of the GumTree algorithm: we use GumTree Spoon\footnote{\url{https://github.com/SpoonLabs/gumtree-spoon-ast-diff}} since such tool delivers the AST {\it diff} nodes in the representation of the Spoon library~\cite{Pawlak2015}, based on a well-designed meta-model for representing Java programs. Therefore, we can analyze the edit scripts returned by GumTree with Spoon.

\noindent\textit{Analysis of the AST diff:} With the AST {\it diff} retrieved, the AST nodes are analyzed to detect the repair patterns. \ToolName~contains a set of detectors, one for each pattern group, because each pattern group has its own definition, which lead us to define a specific strategy for the detection of each of them\footnote{\ToolName~was designed in a modularized way that makes possible the addition of a new pattern detection by extending an existing class and implementing a strategy for the detection of the new pattern.}; the strategies are mainly based on searching and checking code elements/structures in the AST {\it diff}. However, all detectors follow the same general process: it analyzes edit scripts using Spoon based on the defined strategy to detect the pattern it was designed for. Thus, we choose one pattern, \textit{Missing Null-Check}, to be used as an example to describe in details how \ToolName~performs automatic detection. Given the edit scripts from a patch, the strategy of the \textit{Missing Null-Check} detector is the following:

\begin{enumerate}[leftmargin=1.2em]
\item It searches for the addition of a binary operator where one of the two elements is \texttt{null}, i.e., a null-check;

\item It extracts from the null-check the variable being checked (\texttt{variable <operator> null}) or the variable being used to call a method where its return is being checked (\texttt{variable.methodCall() <operator> null});

\item It verifies if the extracted variable is new, i.e., was added in the patch: a) if the variable is not new, a missing null-check was found; b) if the variable is new, it verifies if the new null-check wraps existing code: if it does, a missing null-check was found.
\end{enumerate}

Consider the {\it diff} in Listing~\ref{code:Chart-14}. In the buggy version of this code, in the old line 2166, a null pointer exception had been thrown when the variable \texttt{markers} was null and accessed for a method call. In the fixed version, a conditional block was added to check whether \texttt{markers} is null, and in such case, the method returns, so the program execution does not reach the point of the exception. The added null-check is an instance of the pattern \textit{Missing Null-Check}. Note that the null-check was added in a new conditional block, so this patch also contains an instance of the pattern \textit{Conditional Block Addition with Return Statement}. Additionally, this conditional block was added in four different locations on the code (see \href{http://program-repair.org/defects4j-dissection/\#!/bug/Chart/14}{Chart-14}), which consists in the \textit{Copy/Paste} pattern.

\begin{lstlisting}[caption={Patch for bug \href{http://program-repair.org/defects4j-dissection/\#!/bug/Chart/14}{Chart-14}.},label={code:Chart-14}]
2166      + if (markers == null) {
2167      +    return false;
2168      + }
2166 2169   boolean removed = markers.remove(marker); 
\end{lstlisting}
\vspace*{-10pt}

A missing null-check can appear in different variants beyond the addition of an entire conditional block. In Listing~\ref{code:Chart-26}, for instance, the missing null-check was added in a new conditional by wrapping an existing block of code. This type of change consists of the pattern \textit{Wraps-with if}. Different from \textit{Conditional Block Addition}, the body of the conditional contains existing code in \textit{Wraps-with if}.

\begin{lstlisting}[caption={Patch for bug \href{http://program-repair.org/defects4j-dissection/\#!/bug/Chart/26}{Chart-26}.},label={code:Chart-26}]
1191 1191   ChartRenderingInfo owner = plotState.getOwner();
     1192 + if (owner != null) {
1192 1193      EntityCollection entities = owner.getEntityCollection();
1193 1194      if (entities != null) {
1194 1195         entities.add(new AxisLabelEntity(this, hotspot, 
1195 1196                 this.labelToolTip, this.labelURL));
1196 1197      }
     1198 + }
\end{lstlisting}
\vspace*{-12pt}

Since our detector searches for binary operators involving null-check, it also detects missing null-checks in other structures beyond \texttt{if} conditionals. In Listing~\ref{code:Mockito-29}, for instance, there is an example of a conditional using the ternary operator. When the ternary operator is used, and an existing expression is placed in the \texttt{then} or \texttt{else} expression, we have the pattern \textit{Wraps-with if-else}. Note that this patch is also an instance of the pattern \textit{Single Line} since only one line was affected by the patch.

\begin{lstlisting}[caption={Patch for bug \href{http://program-repair.org/defects4j-dissection/\#!/bug/Mockito/29}{Mockito-29}.},label={code:Mockito-29}]
29    - description.appendText(wanted.toString());
   29 + description.appendText(wanted == null ? "null" : wanted.toString()); 
\end{lstlisting}
\vspace*{-12pt}

For these three example patches, \ToolName~was able to detect all the existing patterns in them, according to the taxonomy presented in Section~\ref{sec:background}.

\vspace*{-4pt}
\section{Evaluation}\label{sec:evaluation}
\vspace*{-2pt}

\noindent\textbf{Method.} Our evaluation consists of running \ToolName~on real patches to measure its ability at detecting the 25 repair patterns.

\vspace*{-2pt}
\noindent\textit{Subject Dataset.} The patches used as input to \ToolName~are from Defects4J~\cite{Just2014}, which consists of 395 patches from six real-world projects (e.g. Apache Commons Lang and Mockito Testing Framework). We chose this dataset since it contains real bugs and all its patches have been annotated with repair patterns~\cite{Sobreira2018defects4J-dissection}, allowing the direct comparison between results generated by \ToolName~and the previous manual detection.

\vspace*{-2pt}
\noindent\textit{Result analysis.} We analyzed the results in two steps. First, we calculated the precision and recall of the \ToolName~for each pattern, using the available manual detection \cite{Sobreira2018defects4J-dissection} as an oracle. We refer to such manual detection as \textit{human detection}, while we refer to the detection produced by \ToolName~as \textit{automatic detection}. Second, we performed manual analysis on the disagreements between the automatic and human detection. For each pattern, two different authors of this paper analyzed all patches where there were disagreements and determined whether \ToolName~actually missed or wrongly detected such pattern. We annotated the disagreements with one of the five diagnostics presented in the first column of Table~\ref{tab:reasons}. Then, we calculated the actual precision and recall for each pattern, using the following formulas: $TP = A + B + DC$, $precision = \frac{TP}{TP + DW}$, $recall = \frac{TP}{TP + HC}$, where $A$ is the number of agreements between \ToolName~and human detection, $B$ is the disagreements when both \ToolName~and human detection may be accepted, $DC$ is the disagreements when \ToolName~detection is correct and $DW$ when \ToolName~detection is wrong, and $HC$ is the disagreements when the human detection is correct.

\noindent\textbf{Results.} The evaluation results are presented in Table~\ref{tab:evaluation-results}: for each pattern, this table shows the precision and recall before (column ``prior'') and after (column ``post'') the disagreement analysis. 

\begin{table}[t]
  \caption{\ToolName~performance.}
  \label{tab:evaluation-results}
  %\small
  \footnotesize
  \centering
  \begin{tabular}{ l l r r r r }
    \toprule
    \multirow{3}{*}{Pattern} & \multirow{3}{*}{Variant} & \multicolumn{2}{c}{Prior} & \multicolumn{2}{c}{Post} \\
    {} & {} & \multicolumn{1}{c}{Precision} & \multicolumn{1}{c}{Recall} & \multicolumn{1}{c}{Precision} & \multicolumn{1}{c}{Recall} \\
    {} & {} & \multicolumn{1}{c}{(\%)} & \multicolumn{1}{c}{(\%)} & \multicolumn{1}{c}{(\%)} & \multicolumn{1}{c}{(\%)} \\
    \midrule
    
    \rowcolor{gray!15}
    {} & Addition & 74.75 & 93.67 & 99.00 & 98.02 \\
    \rowcolor{gray!15}
    {} & ~~~~~~$''$~~~~~~with Return Statement & 90.12 & 94.81 & 100.00 & 96.47 \\
    \rowcolor{gray!15}
    {} & ~~~~~~$''$~~~~~~with Exception Throwing & 93.75 & 90.91 & 96.88 & 91.18 \\
    \rowcolor{gray!15}
    \multirow{-4}{*}{Conditional Block} & Removal & 60.71 & 77.27 & 86.67 & 89.66 \\
    
    {} & Logic Modification & 82.22 & 75.51 & 91.11 & 83.67 \\
    {} & ~~~~$''$~~~~ Expansion & 90.20 & 95.83 & 92.16 & 97.92 \\
    {} & ~~~~$''$~~~~ Reduction & 76.92 & 83.33 & 76.92 & 100.00 \\
    \multirow{-4}{*}{Expression Fix} & Arithmetic Fix & 69.57 & 50.00 & 91.67 & 64.71 \\
    
    \rowcolor{gray!15}
    {} & Wraps-with if & 74.19 & 95.83 & 83.87 & 96.30 \\
    \rowcolor{gray!15}
    {} & ~~~~~~~~~$''$~~~~~~~~~if-else & 81.25 & 84.78 & 92.00 & 90.20 \\
    \rowcolor{gray!15}
    {} & ~~~~~~~~~$''$~~~~~~~~~else & 16.67 & 100.00 & 33.33 & 100.00 \\
    \rowcolor{gray!15}
    {} & ~~~~~~~~~$''$~~~~~~~~~try-catch & 100.00 & 100.00 & 100.00 & 100.00 \\
    \rowcolor{gray!15}
    {} & ~~~~~~~~~$''$~~~~~~~~~method & 78.57 & 78.57 & 85.71 & 85.71 \\
    \rowcolor{gray!15}
    {} & ~~~~~~~~~$''$~~~~~~~~~loop & 40.00 & 100.00 & 60.00 & 100.00 \\
    \rowcolor{gray!15}
    {} & Unwraps-from if-else & 42.11 & 61.54 & 57.89 & 68.75 \\
    \rowcolor{gray!15}
    {} & ~~~~~~~~~~~$''$~~~~~~~~~~~~try-catch & 100.00 & 100.00 & 100.00 & 100.00 \\
    \rowcolor{gray!15}
    \multirow{-9}{*}{Wraps/Unwraps} & ~~~~~~~~~~~$''$~~~~~~~~~~~~method & 45.45 & 83.33 & 54.55 & 85.71 \\
    
    Single Line & -- & 100.00 & 97.96 & 100.00 & 100.00 \\
    
    \rowcolor{gray!15}
    {} & Variable & 66.67 & 76.19 & 82.35 & 89.36 \\
    \rowcolor{gray!15}
    \multirow{-2}{*}{Wrong Reference} & Method & 68.42 & 83.87 & 86.84 & 89.19 \\
    
    {} & Positive & 95.45 & 84.00 & 100.00 & 100.00 \\
    \multirow{-2}{*}{Missing Null-Check} & Negative & 96.67 & 90.63 & 100.00 & 96.77 \\
    
    \rowcolor{gray!15}
    Copy/Paste & -- & 56.16 & 85.42 & 91.78 & 90.54 \\
    
    Constant Change & -- & 77.27 & 89.47 & 90.91 & 90.91 \\
    
    \rowcolor{gray!15}
    Code Moving & -- & 60.00 & 85.71 & 81.82 & 100.00 \\
    
    \midrule
    Overall &  & 78.26 & 86.95 & 91.53 & 92.39 \\
    \bottomrule
  \end{tabular}
  \vspace*{-20pt}
\end{table}

We observed that \ToolName~has a high overall precision and recall, even when just comparing it directly with the human detection (see the last line in the table). For the most recurring pattern group, \textit{Conditional Block}, both detections agreed on 194 instances of such pattern (prior). After the disagreement analysis, we found that \ToolName~detected 39 new instances of such pattern, which increased the precision and recall of the \ToolName~(post), for at least 86\% and 89\%, respectively.

For some less recurring patterns, \textit{Single Line} and \textit{Missing Null-Check}, \ToolName~performed well by detecting 96 and 50 instances of these patterns in agreement with the human detection, respectively. In fact, for \textit{Single Line}, the only two instances missing by \ToolName~were not truly instances of such pattern.

However, we identified some particular patterns that \ToolName~did not perform well. \ToolName~found 95 instances of the patterns from the group \textit{Wraps/Unwraps} in agreement with the human detection. On the disagreement analysis, we identified that \ToolName~detected 7 new instances of this group, but that also generated 30 false positives. The major responsible for these false positives are the pattern variants involving \texttt{if}, \texttt{else} and \texttt{method}.

\begin{table}[t]
  \caption{Overall absolute results on the disagreement analysis and reasons for automatic detection differing from the manual detection.}
  \label{tab:reasons}
  %\small
  \footnotesize
  \centering
  \begin{tabular}{ l r r }
    \toprule
    Diagnostic & \# Occurrences & Related Reason \\
    \midrule
    DW (\ToolName~false positive) & 73 & \#1, \#7 \\
    DC (\ToolName~true positive) & 90 & \#1, \#4 \\
    HW (human detection false positive) & 24 & \#5, \#6 \\
    HC (human detection true positive) & 65 & \#2, \#7 \\
    B (both could be accepted) & 33 & \#1, \#3 \\
    \midrule
    A (agreements) & 666 & {} \\
    TP (correct detection = A + B + DC) & 789 & {} \\
    \bottomrule
  \end{tabular}
  \vspace*{-1pt}
\end{table}

\noindent\textbf{Discussion.} During the disagreement analysis, we also investigated why \ToolName~failed or differed from the human analysis. Table~\ref{tab:reasons} relates the diagnostics with the reasons for the disagreements, which we discuss as follows.

\noindent Reason \#1: Global human vision versus AST-based analysis. The GumTree algorithm identifies implicit structures that are not visible by  humans. For instance, in \href{http://program-repair.org/defects4j-dissection/\#!/bug/Mockito/18}{Mockito-18}, both automatic and manual detections found the pattern \textit{Conditional Block Addition with Return Statement}. However, the automatic detection also found the pattern \textit{Wraps-with if-else}. In this patch, the human sees the structure as in Listing~\ref{code:Mockito-18a}, while the structure considered by \ToolName~is like in Listing~\ref{code:Mockito-18b}. In other words, the new conditional block wraps a part of the code, but with an implicit block. On these occurrences, we considered that both automatic and manual detection could be accepted.

\noindent 
\begin{minipage}[b]{0.5\textwidth}
  \begin{lstlisting}[caption={Human vision.},label={code:Mockito-18a},linewidth=0.96\columnwidth,xleftmargin=5pt]
  + } else if (type == Iterable.class) {            
  +    return new ArrayList<Object>(0);            
    } else if (type == Collection.class) {
    [...]
  \end{lstlisting}
\end{minipage}
\begin{minipage}[b]{0.5\textwidth}
\begin{lstlisting}[caption={AST-based analysis.},label={code:Mockito-18b},linewidth=0.96\columnwidth,xleftmargin=0pt]
+ } else {
+    if (type == Iterable.class) {            
+       return new ArrayList<Object>(0);            
     } else {
        if (type == Collection.class) {
  [...]
\end{lstlisting}
\end{minipage}

\vspace*{-15pt}

Still on the global human vision versus AST-based analysis discussion, due to fine-grained changes, \ToolName~takes into account small changes that do not make sense as the composition of a pattern in some cases. For instance, \ToolName~detected the \textit{Copy/Paste} pattern in \href{http://program-repair.org/defects4j-dissection/\#!/bug/Chart/3}{Chart-3}. Even though the two additions have a high similarity, these changes are not enough to be considered as an instance of the pattern \textit{Copy/Paste}, so we determined this as a false positive generated by \ToolName.

In the same direction, \ToolName~takes into account \textit{relevant} small changes that  humans may not identify in big patches.
In these big patches,  humans may intuitively consider only the global vision of the patch and miss smaller changes.
For instance, \href{http://program-repair.org/defects4j-dissection/\#!/bug/Math/64}{Math-64} has several changes: one of them is the addition of a block with three lines of code in two different locations (i.e., \textit{Copy/Paste}), which was missed by  human detection.

\noindent Reason \#2: The automatic detection relies on rules defined by humans (i.e., the authors of this paper), and it is difficult to identify all cases where an instance of a pattern may exist, thus \ToolName~missed some pattern instances. The \textit{Expression Fix} detector is the primary responsible for these missing detections. For instance, it missed the detection of an arithmetic expression fix in \href{http://program-repair.org/defects4j-dissection/\#!/bug/Math/77}{Math-77}, where an arithmetic operation occurs with the assignment operator \texttt{+=}, which was replaced by a non-arithmetic assignment operator.

\noindent Reason \#3: There are some borderline cases where a given pattern may fit or not. For instance, in line 1167 of \href{http://program-repair.org/defects4j-dissection/\#!/bug/Time/17}{Time-17}, one could consider the removed method call as a part of the arithmetic expression used as argument for such method call, and another one could not. Only the manual detection detects such statement as an instance of the \textit{Arithmetic Expression Fix} pattern, but we considered that detecting it or not can be both accepted.

\noindent Reason \#4: The automatic detection applies the same rules for all the patches while it is a difficult task to be done by humans. Therefore, some pattern instances were missed by the manual detection due inconsistencies between patches.

\noindent Reason \#5: In the manual analysis,  humans may consider the semantic of the changes (even without noticing) and make assumptions on how the developer could write a patch that matches one of the patterns. For instance, \href{http://program-repair.org/defects4j-dissection/\#!/bug/Mockito/28}{Mockito-28} had been considered as an instance of the \textit{Single Line} pattern. Semantically, it could be correct, but such pattern should be limited to changes affecting a single line or a single statement in a given patch.

\noindent Reason \#6: A misconception of the patch can impact the human analysis. For instance, in \href{http://program-repair.org/defects4j-dissection/\#!/bug/Lang/50}{Lang-50}, the manual detection found an instance of the pattern \textit{Logic Expression Modification} in line 285 (and also in the new line 463, which is the same case, i.e. \textit{Copy/Paste}). However, the existing conditional block in line 285 was actually completely changed: the statement inside it was unwrapped in the patch, and the conditional was deleted. Then, an existing conditional block in the code took the place of the conditional considered as having its logic expression modified, i.e. a moving happened, not characterizing a genuine logic expression modification.

\noindent Reason \#7: GumTree is a sophisticated algorithm that may return imprecise results for some patches. For example, it can consider the change over some elements when it is not the case. As a consequence, PPD incorrectly detects or misses some pattern instances.

\section{Threats to Validity}\label{sec:threats}
\vspace*{-2pt}

\noindent\textit{Internal validity.} The ultimate precision and recall calculated when evaluating the performance of the \ToolName~are based on a manual disagreement analysis. This analysis can be subject to small errors and misconception, typical of any manual work. To mitigate this, such analysis was performed to each pattern group by two authors of this paper, in live discussion sessions.

\noindent\textit{External validity.} We have evaluated \ToolName~on patches from Defects4J. However, since Defects4J may not be representative on all different cases on fixing bugs using one of the 25 patterns, it is possible that \ToolName~still cannot generalize for systems including patches that differ a lot from those in Defects4J. Moreover, detected repair patterns are Java-based, therefore our detector is limited to systems written in this language.

\section{Related Work}\label{sec:related-works}
\vspace*{-2pt}

Martinez et al.~\cite{Martinez2013} also reported on the automatic detection of bug fix patterns at the AST level. The main differences between their work and our work are the following.
First, they focused on 18 bug fix patterns from~\cite{Pan2009} while we focused on 25 patterns from~\cite{Sobreira2018defects4J-dissection}.
Second, they used the ChangeDistiller AST differencing algorithm~\cite{Fluri2007} while we use GumTree~\cite{Falleri2014}. The latter outperforms the former by maximizing the number of AST node mappings, minimizing the edit script size, and detecting better move actions~\cite{Falleri2014}. Moreover, they pointed out that ChangeDistiller works at the statement level, preventing the detection of certain fine-grain patterns.
Third, they formalized a representation for change patterns and used this representation to specify patterns. Then, to detect a pattern, a match of its specification must happen in a given edit script. However, such representation is based on change type (e.g. addition) over code elements (e.g. \texttt{if}), which does not support the specification of patterns such as \textit{Single Line}.

%\noindent\textit{Detecting repair patterns.} Soto et al. \cite{Soto2016} aimed at identifying how many patches contain the repair patterns presented by \cite{Kim2013par}. They analyzed 4M bug fix commits and found that less than 15\% of them contain one of those patterns. Different from our work, they analyzed bug fixes by processing pre- and post-fix files separately, which results in approximated pattern counters.

%\noindent\textit{Mining repair actions.} Martinez and Monperrus \cite{Martinez2015}

%\noindent\textit{Analysis on Defects4J patches.} Besides Sobreira et al. \cite{Sobreira2018defects4J-dissection}, Motwani et al. \cite{Motwani2018} also manually annotated Defects4J with characteristics on the patches. The authors of those two works were concerned with the analysis and annotation of Defects4J to achieve their goals, which is different from our work towards a general approach to automate the extraction of patches' characteristics.

\section{Final Remarks}\label{sec:conclusion}

In this paper, we report on \ToolName, a detector of repair patterns in bug fixes. Through an evaluation on Defects4J, we found that \ToolName~has a good performance in general, and for some patterns (e.g., \textit{Missing Null-Check}) it can even perform better than human detection. Moreover, a fruit of the disagreement analysis, we found that human detection made fewer mistakes (24) than \ToolName~(73), but also detected less exclusive occurrences (65) than \ToolName~(90). As future work, we intend to conduct experiments over other bug datasets to evaluate the scalability of \ToolName~and also to compare bug datasets, which may guide researchers on automatic program repair at choosing datasets when evaluating their tools. Finally, we intend to create a visualization for patches where the repair patterns are highlighted, to support the human patch comprehension task.

\noindent\textbf{Tool Availability.} \ToolName~is part of the project ADD, which is publicly available at:

\begin{center}
\url{https://github.com/lascam-UFU/automatic-diff-dissection}
\end{center}

\noindent One can find instructions in such repository on how to use \ToolName~and also to reproduce the results on Defects4J presented in our evaluation (Section~\ref{sec:evaluation}).

\bibliographystyle{sbc}
\bibliography{references}

\begin{thebibliography}{}

\bibitem[Durieux et~al. 2017]{Durieux2017-npefix}
Durieux, T., Cornu, B., Seinturier, L., and Monperrus, M. (2017).
\newblock {Dynamic Patch Generation for Null Pointer Exceptions Using
  Metaprogramming}.
\newblock In {\em SANER '17}.

\bibitem[Falleri et~al. 2014]{Falleri2014}
Falleri, J.-R., Morandat, F., Blanc, X., Martinez, M., and Monperrus, M.
  (2014).
\newblock {Fine-grained and Accurate Source Code Differencing}.
\newblock In {\em ASE '14}, pages 313--324.

\bibitem[Fluri et~al. 2007]{Fluri2007}
Fluri, B., Wuersch, M., PInzger, M., and Gall, H. (2007).
\newblock {Change Distilling: Tree Differencing for Fine-Grained Source Code
  Change Extraction}.
\newblock {\em TSE}, 33(11):725--743.

\bibitem[Just et~al. 2014]{Just2014}
Just, R., Jalali, D., and Ernst, M.~D. (2014).
\newblock {Defects4J: A Database of Existing Faults to Enable Controlled
  Testing Studies for Java Programs}.
\newblock In {\em ISSTA '14}, pages 437--440.

\bibitem[Martinez et~al. 2013]{Martinez2013}
Martinez, M., Duchien, L., and Monperrus, M. (2013).
\newblock {Automatically Extracting Instances of Code Change Patterns with AST
  Analysis}.
\newblock In {\em ICSM '13}, pages 388--391.

\bibitem[Monperrus 2018]{Monperrus2018}
Monperrus, M. (2018).
\newblock {Automatic Software Repair: a Bibliography}.
\newblock {\em ACM Computing Surveys}, 51(1):17:1--17:24.

\bibitem[Pan et~al. 2009]{Pan2009}
Pan, K., Kim, S., and Whitehead, Jr., E.~J. (2009).
\newblock {Toward an understanding of bug fix patterns}.
\newblock {\em EmSE}, 14(3):286--315.

\bibitem[Pawlak et~al. 2015]{Pawlak2015}
Pawlak, R., Monperrus, M., Petitprez, N., Noguera, C., and Seinturier, L.
  (2015).
\newblock {Spoon: A Library for Implementing Analyses and Transformations of
  Java Source Code}.
\newblock {\em {Software: Practice and Experience}}, 46:1155--1179.

\bibitem[Sobreira et~al. 2018]{Sobreira2018defects4J-dissection}
Sobreira, V., Durieux, T., Madeiral, F., Monperrus, M., and Maia, M.~A. (2018).
\newblock {Dissection of a Bug Dataset: Anatomy of 395 Patches from Defects4J}.
\newblock In {\em SANER '18}.

\end{thebibliography}

\end{document}